\documentclass[journal]{IEEEtran}

\usepackage{graphicx}
\usepackage{array}
\makeatletter
\providecommand{\tabularnewline}{\\}
\makeatother

\usepackage[ruled]{algorithm2e} 

\usepackage[export]{adjustbox}
\usepackage[caption=false]{subfig}
\usepackage[T1]{fontenc}  
\usepackage{textcomp} 
\usepackage{mathtools}
\usepackage{amsmath}
\usepackage{amsfonts}
\usepackage{array}
\usepackage{multirow}
\usepackage{balance}
\allowdisplaybreaks

\usepackage[normalem]{ulem}
\usepackage{pdfcomment}
\usepackage{xcolor}
\definecolor{red}{rgb}{1,0,0}

\begin{document}

\title{Optimizing city-scale traffic through modeling observations of vehicle movements}

\author{Fan~Yang,
        Alina~Vereshchaka,
        Bruno~Lepri,
        Wen~Dong,~\IEEEmembership{member,~IEEE}%
\thanks{F. Yang was with the Department
of Computer Science and Engineering, University at Buffalo, Buffalo,
NY, 14260 USA e-mail: (fyang24@buffalo.edu).}%
\thanks{A. Vereshchaka was with the Department
of Computer Science and Engineering, University at Buffalo, Buffalo,
NY, 14260 USA e-mail: (avereshc@buffalo.edu).}%
\thanks{B. Lepri is with the Mobile and Social Computing Laboratory, Bruno
Kessler Foundation, 38122 Trento, Italy (e-mail: lepri@fbk.eu).}%
\thanks{W. Dong was with the Department
of Computer Science and Engineering, University at Buffalo, Buffalo,
NY, 14260 USA e-mail: (wendong@buffalo.edu).}%
\thanks{Manuscript received Month date, year; Month date, year}}
\maketitle

\begin{abstract}
The capability of traffic-information systems to sense the movement of millions of users and offer trip plans through mobile phones has enabled a new way of optimizing city traffic dynamics, turning transportation big data into insights and actions in a closed-loop and evaluating this approach in the real world. Existing research has applied dynamic Bayesian networks and deep neural networks to make traffic predictions from floating car data, utilized dynamic programming and simulation approaches to identify how people normally travel with dynamic traffic assignment for policy research, and introduced Markov decision processes and reinforcement learning to optimally control traffic signals. However, none of these works utilized floating car data to suggest departure times and route choices in order to optimize city traffic dynamics. In this paper, we present a study showing that floating car data can lead to lower average trip time, higher on-time arrival ratio, and higher Charypar-Nagel score compared with how people normally travel. The study is based on optimizing a partially observable discrete-time decision process and is evaluated in one synthesized scenario, one partly synthesized scenario, and three real-world scenarios. This study points to the potential of a ``living lab'' approach where we learn, predict, and optimize behaviors in the real world.
\end{abstract}

\section{Introduction}

With 80\% of newly sold vehicles in the U.S. able to communicate vehicle state through a telematics unit, and 57\% of the population connected to the Internet by smart phones, datasets that track vehicles are increasingly available for transportation and policy researchers to study human mobility at scale. Such datasets contain rich information --- from how drivers plan their daily activities and trips at the microscopic level to how road networks respond to transportation demand at the macroscopic level \cite{an2011agent,kung2014,zheng2017}. Traffic-information providers such as Google Traffic and INRIX can play an essential role in city traffic dynamics by suggesting optimal trip plans according to the observed movements of millions of vehicles. At the same time, artificial intelligence is accelerating workplace transition and the way people travel at a pace forecasting-based policy research might ultimately be unable to keep up with. This trend demands a paradigm that leverages traffic big data to deliver agile, quantifiable, and scalable solutions to our real-world transportation problems. Algorithms based on graphical models \cite{horvitz2005,yang2019optimal} and neural networks \cite{lv2015traffic,polson2017deep,ma2017learning,zhang2017deep,zhao2017lstm} have been developed to make traffic predictions from the movement of millions of vehicles, but none of them utilizes these predictions to optimize complex city traffic dynamics. Similarly, optimization algorithms have been developed to identify how people normally travel through traffic assignment \cite{ziliaskopoulos2000linear,zolfpour2014modeling,smith1995transims,horni2016multi} in policy research and to optimize traffic-light schedules \cite{timotheou2015distributed,lin2012efficient,genders2016using,casas2017deep,el2013multiagent}, but none directly connects the observed vehicle trajectories with suggested driver departure times and route choices in a closed-loop control to optimize city traffic.

In this paper, we present a simulation study showing that by transforming the observed probe-vehicle movement data into traffic predictions and suggestions about optimal departure times and route choices in closed-loop control, we can achieve lower average trip time and higher on-time arrival ratio, and thus a higher Charypar-Nagel score \cite{horni2016multi}, compared with how people normally travel. To achieve this, we model the traffic optimization problem as a partially observable Markov decision process, where the observations are the numbers of ``probe'' vehicles at road links and buildings, the future expected reward to optimize is the Charypar-Nagel scoring function \cite{horni2016multi}, the control variables are related to departure times and route choices, and the dynamics are modeled as a queuing network approximated with a discrete-event model with neural network components. The simulation study is conducted using one synthesized scenario \cite{horni2016multi}, one partly real and partly synthesized scenario \cite{ziemke2015integrating}, and three real-world scenarios \cite{kickhofer2016creating,de2014d4d}.

The uniqueness of this paper is that we combine machine learning methods and big floating car data to prototype a new traffic optimization approach. The variational tracking and optimal control algorithms that we developed \cite{yang2019optimal,yang2020variational,yang2020optimal} are complex and less straightforward, so we specifically develop a new particle filter algorithm to track and optimize traffic by extending our previous work \cite{yang2019optimal,yang2018predicting}. We also provide a detailed evaluation of this approach to traffic prediction and control in synthesized, partly real and partly synthesized, and real-world scenarios. Our approach not only simulates traffic jams during rush hours but also {\em predicts} from the trajectories of probe vehicles whether today's traffic jams will be formed earlier or last longer than usual, and helps drivers to {\em decide} and {\em plan} how to use the road network more efficiently. 

\begin{figure}
\hfil\includegraphics[width=1.0\linewidth]{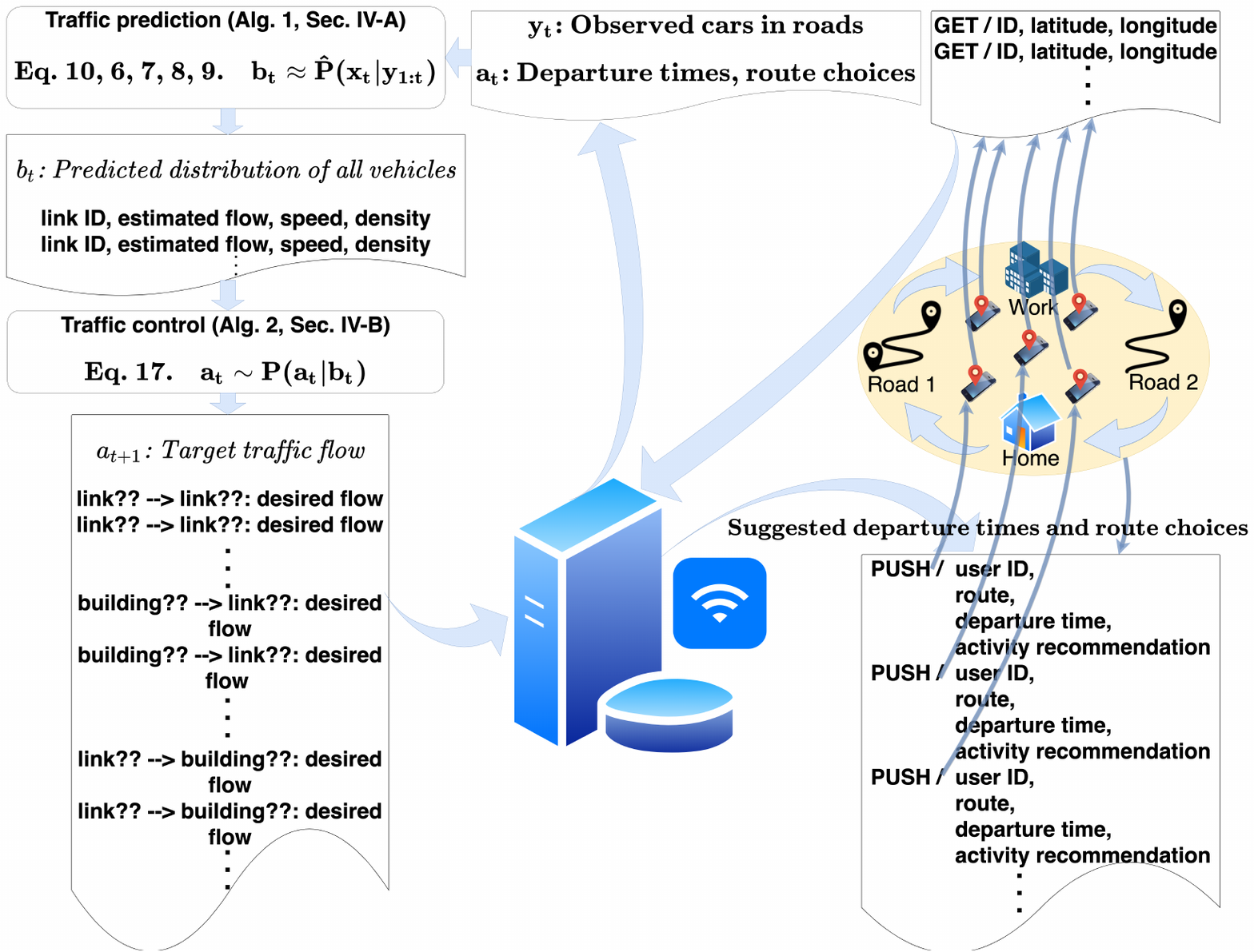} \hfil\vspace{-.1in}
\caption{\label{fig:workflow} A use-case diagram of traffic monitoring and control with drivers' navigation apps at high level.}\vspace{-.1in}
\end{figure}

A use-case diagram in Fig. \ref{fig:workflow} shows a high-level view of how the theory in this paper can be deployed in the real world to track traffic state and optimize traffic dynamics. Floating vehicle locations from drivers' navigation apps are aggregated into noisy observations about traffic demands and state ($y_t$ and $a_{t-1}$). These noisy observations are fed into the traffic prediction algorithm (Algorithm \ref{alg:PF}, Sec. \ref{subsec:traffic_prediction}) to calculate the improved traffic state estimation as $b_t\approx\hat P(x_t|y_{1:t})$ using Eq. \ref{eq:belief}, which in turn uses Eqs. \ref{eq:pf-v}, \ref{eq:pf-x}, \ref{eq:pf-i}, and \ref{eq:pf-path}. The improvement comes from aggregating past observations $y_{1:t}$, and using a Bayesian formulation $\hat P(x_t|y_{1:t})$. The improved traffic state estimation is then fed into the traffic control algorithm (Algorithm \ref{alg:control}, Sec. \ref{subsec:traffic_control}) to calculate the desired/target traffic flow as $a_t\sim P(a_t|b_t)$ using Eq. \ref{eq:M}. The desired traffic flow control is in the form of what proportions of traffic should be directed to downstream links/facilities and is used to suggest departure times and route choices through drivers' navigation apps. This improves traffic with better and real-time information $b_t$ about traffic state.

The remainder of this paper is organized as follows. 
Section \ref{literature} discusses other research efforts on making predictions and identifying optimal controls in the road transportation network from noisy observations.
Section \ref{problem} introduces the discrete-event decision process to define the problem of optimizing  city traffic dynamics. Section \ref{sec:methodology} details a particle filter algorithm that predicts and optimizes traffic from noisy observations. Section \ref{experiments} evaluates the performance of a discrete-event decision process and particle filter against other model-based and model-free methods on synthesized and real-world datasets. In Section \ref{conclusions} we discuss the implications and limitations of our work and draw some conclusions.

\section{Related Works}

\label{literature} In this paper, we apply a discrete-event decision process \cite{yang2019optimal,yang2020variational,yang2020optimal} to predict complex city traffic dynamics from the trajectories of probe vehicles and accordingly optimize that traffic through departure times and route choices. Related research in intelligent transportation systems falls into two research streams: traffic prediction and traffic optimization.

Traditionally, traffic prediction was based on traffic cameras, inductive-loop traffic detectors, and similar technologies installed at fixed locations to capture speed, flow, and density. Algorithms include extended Kalman filter \cite{wang2006renaissance}, localized extended Kalman filter \cite{van2012localized}, extended generalized Treiber-Helbing filter \cite{van2010robust}, and particle filter \cite{xie2018generic}. More recently, mobile phones and the Internet of Things have provided a new way to log the trajectories of millions of vehicles. These trajectories introduce an unprecedented opportunity to estimate home and work locations \cite{kung2014}, identify trip purposes and special events \cite{calabrese2011}, model driver behaviors \cite{junior2017}, and track road network dynamics \cite{zheng2017,zhang2017deep}. Algorithms to leverage these trajectory data include probabilistic Bayesian networks \cite{horvitz2005}, deep neural networks \cite{lv2015traffic,polson2017deep}, convolutional neural networks \cite{ma2017learning,zhang2017deep}, graph convolutional networks \cite{yu2017spatio,li2017graph}, and recurrent neural networks \cite{zhao2017lstm}. Traffic prediction has been used for model-calibration in policy research, controlling traffic signals, and informing drivers. However, to the best of our knowledge the research in this paper is the first to optimize traffic by suggesting optimal departure times and route choices. It is also new to apply an agent-based \cite{an2011agent} discrete-event model to both predict traffic and visualize how vehicles move in a city-scale road network in accordance with where probe vehicles move and how traffic policies change vehicle behavior.

Traffic optimization is conducted for transportation forecasting and traffic control. Transportation forecasting is a four-step process --- trip generation, trip distribution, mode choice, and traffic assignment --- that estimates the future usage of specific transportation facilities in order to assist policy research, such as assessing land-development impact \cite{zhu2018integrating}. Traffic optimization is conducted in the fourth steps through either mathematical programming \cite{ziliaskopoulos2000linear,zolfpour2014modeling} or simulation \cite{smith1995transims,ben2012dynamic,horni2016multi}, with the assumption that people select routes with the minimum travel times at the equilibrium. While optimization is used to identify reasonable routes in traffic assignment, its usage in this paper is to optimize the overall utilities of all people in a transportation network in response to instant traffic prediction based on the observed probe-vehicle locations. Our utility to optimize is the Charypar-Nagel scoring function \cite{horni2016multi}, which entails minimizing not only travel time but also uncertainty in arrival time, as well as the impact of traffic fluctuation on planned activity duration. Various approaches have been applied to optimize traffic-signal control, including mixed-integer linear programming \cite{timotheou2015distributed}, model predictive control \cite{lin2012efficient}, Q-learning \cite{genders2016using}, policy gradient \cite{casas2017deep}, and multi-agent reinforcement learning \cite{el2013multiagent}. Numerous works have used Markov decision processes and reinforcement learning to optimize other aspects of transportation, such as the accessibility of taxicabs \cite{rong2016rich,yu2019markov}. While existing research optimizes transportation dynamics by setting traffic-light schedules, we optimize by advising drivers about ideal departure times and route choices.

As city-scale floating car data is becoming available to the academia \cite{de2014d4d,blondel2015survey}, we previously developed a variational solution to the intractable problem in complex-network optimal control \cite{yang2019optimal,yang2020variational,yang2020optimal}. The essence of the variational solution is to both find optimal trip plans in a tractable surrogate optimization problem and establish a mathematical guarantee that the performance of the solution in the original problem is not worse. In comparison with our previous works, the focus in this paper is developing conceptually straightforward algorithms and providing comprehensive evaluation in the transportation domain.

\section{Problem Statement}\label{problem}

In this Section, we introduce a discrete-event decision process (Eq. \ref{eq:podedp}, Sec. \ref{subsec:dedp}) to define the traffic optimization problem. In doing so, we specify the utility to optimize, the states, the events, the observations, and the control variables (Sec. \ref{subsec:model_traffic_control}).  

Overall, our purpose is to optimally control transportation dynamics in order to a) minimize the total travel time, b) minimize the delay for cars to arrive at their destinations due to traffic fluctuations, and c) minimize the effect of commuting on traffic congestion so that people can perform activities at destinations for an ideal amount of time without incurring significant travel-time increases. Optimal control is achieved by advising individual drivers about downstream links in route planning and the time to leave a given location. The resulting effect is stochastic. The performance indicators of a transportation network are calculated and estimated by the states of all the vehicles and probe vehicles, respectively, where the probe vehicles account for only a small fraction of the vehicle population. Optimal control is also computed from the observed probe-vehicle populations at different locations. This kind of control is applicable where a traffic-information provider (such as Google Maps or a traffic authority) provides trip plans according to the locations reported by a small number of drivers. We assume discrete-event dynamics, where the system state consists of numbers of vehicles at road links and buildings and is driven by elementary events in the form of an individual vehicle moving from one location to the next.

\subsection{Discrete-Event Model for Inference and Decision Making}
\label{subsec:dedp}Here, we introduce a discrete-event model called the \emph{stochastic kinetic model} \cite{gillespie2007stochastic,wilkinson2011stochastic}, which captures the dynamics of a complex social system driven by a set of events. A discrete-event model is a versatile model for describing a wide range of dynamics in various fields. It has many other names, including stochastic kinetic model \cite{gillespie2007stochastic,wilkinson2011stochastic}, stochastic Petri net \cite{goss1998quantitative}, system dynamics model \cite{forrester1969urban}, multi-agent model specified through a flow chart or a state chart \cite{borshchev2013big}, Markov jump process \cite{opper2008variational,rao2013fast}, continuous time Bayesian network \cite{nodelman2002continuous}, and production rule system \cite{newell1972human}. The premise of introducing a discrete-event simulation model \cite{marsan1994modelling,wilkinson2011stochastic,borshchev2013big} to specify road network dynamics is that complex system dynamics can be described by a set of microscopic events that individually bring only minimal changes but in sequence induce complex behaviors. Using a discrete-event model, we specify traffic dynamics in a road network with a set of stochastic events --- a driver starting a trip, moving to the next road, and ending a trip, for example --- and we introduce a set of control variables to influence driver choices in response to the environment.

Let $\mathbb X^{(1)},\cdots,\mathbb X^{(M)}$ denote the individuals belonging to the $M$ species in the system. A discrete-event process is generated by a set of events in the form of a production
\begin{align}
  &\alpha_{v}^{(1)}\mathbb X^{(1)}\!+\!...\!+\!\alpha_{v}^{(M)}\mathbb 
X^{(M)}\!\overset{c_v}\to\!\beta_{v}^{(1)}\mathbb 
X^{(1)}\!+\!...\!+\!\beta_{v}^{(M)}\mathbb X^{(M)}.\label{eq:reaction}
\end{align}
This production is interpreted as having event rate coefficient $c_{v}$ (the probability per unit time as time goes to 0). $\alpha_{v}^{(1)}$ individuals of species $1$, ..., $M$ interact according to event $v$, resulting in their removal from the system, and $\beta_{v}^{(1)}$ individuals of species $1$, ..., $M$ are introduced into the system. Thus, event $v$ changes the populations by $\Delta_{v}=(\beta_{v}^{(1)}- \alpha_{v}^{(1)},\cdots,\beta_{v}^{(M)}-\alpha_{v}^{(M)})$.
Let $x_{t}=(x_{t}^{(1)},\dots,x_{t}^{(M)})$ be the populations of the species in the system at time $t$. A stochastic kinetic process initially in state $x_{0}$ at time $t=0$ can be simulated through the Gillespie algorithm 
\cite{gillespie2007stochastic} by iteratively (1) sampling the time $\tau$ to the next event according to exponential distribution $\tau\sim\mbox{Exponential}(h_{0}(x_t,c))$, where $h_{0}(x,c)=\sum_{v=1}^{V}h_{v}(x_t,c_{v})$ is the rate of all events and $h_v(x_t,c_v)$ is the rate of event $v$, (2) simulating event $v$ according to categorical distribution 
$v\sim(\frac{h_{1}}{h_{0}},\dots,\frac{h_{V}}{h_{0}})$ conditional on event time $\tau$, and (3) updating the system time $t\leftarrow t+\tau$ and populations $x\leftarrow x+\Delta_{v}$ until the termination condition is satisfied. In this algorithm, event rate $h_{v}(x_t,c_{v})$ is the rate constant $c_{v}$ multiplying a total of $\prod_{m=1}^{M}(x^{(m)}_t)^{\alpha_{v}^{(m)}}$ different ways for individuals to interact in the system, assuming homogeneous populations. Exponential distribution is the maximum entropy distribution given the rate constant, and consequently it is most likely to occur in natural reactions \cite{gillespie2007stochastic}.
\begin{figure}
\subfloat[Discrete event process]{
\includegraphics[width=0.49\linewidth]{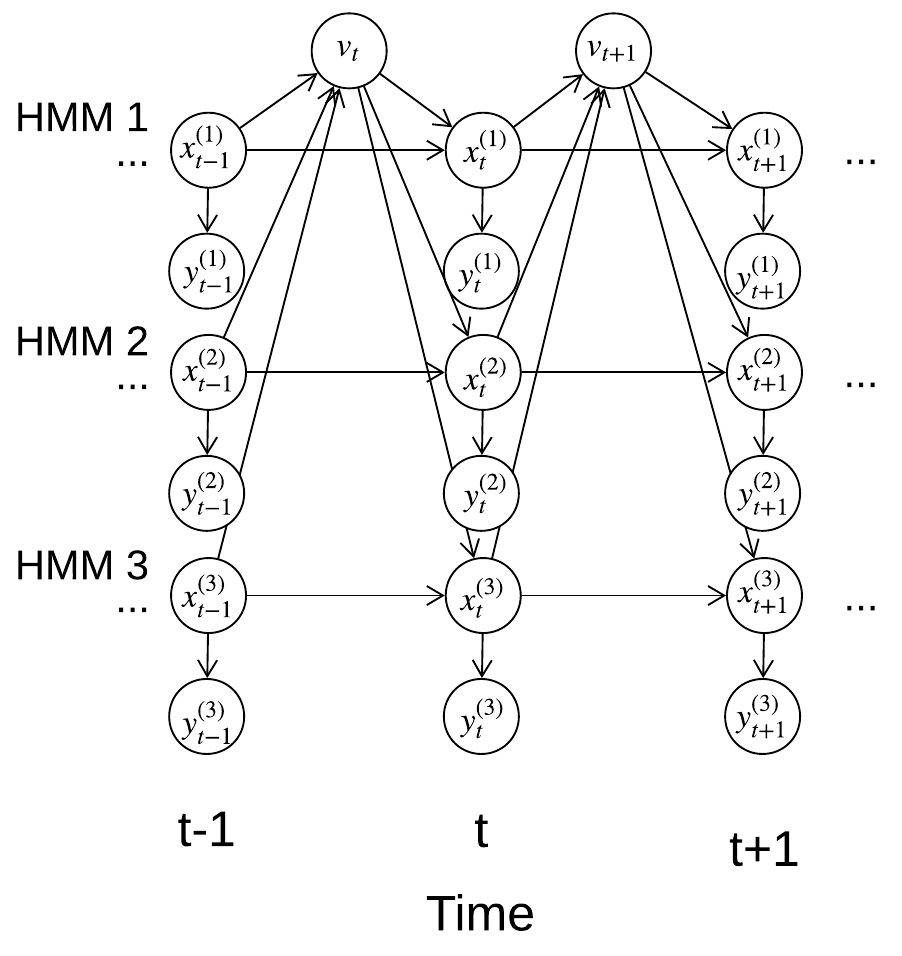} 
\label{fig:SKM}}
\subfloat[Discrete event decision process]{\centering
\includegraphics[width=0.49\linewidth]{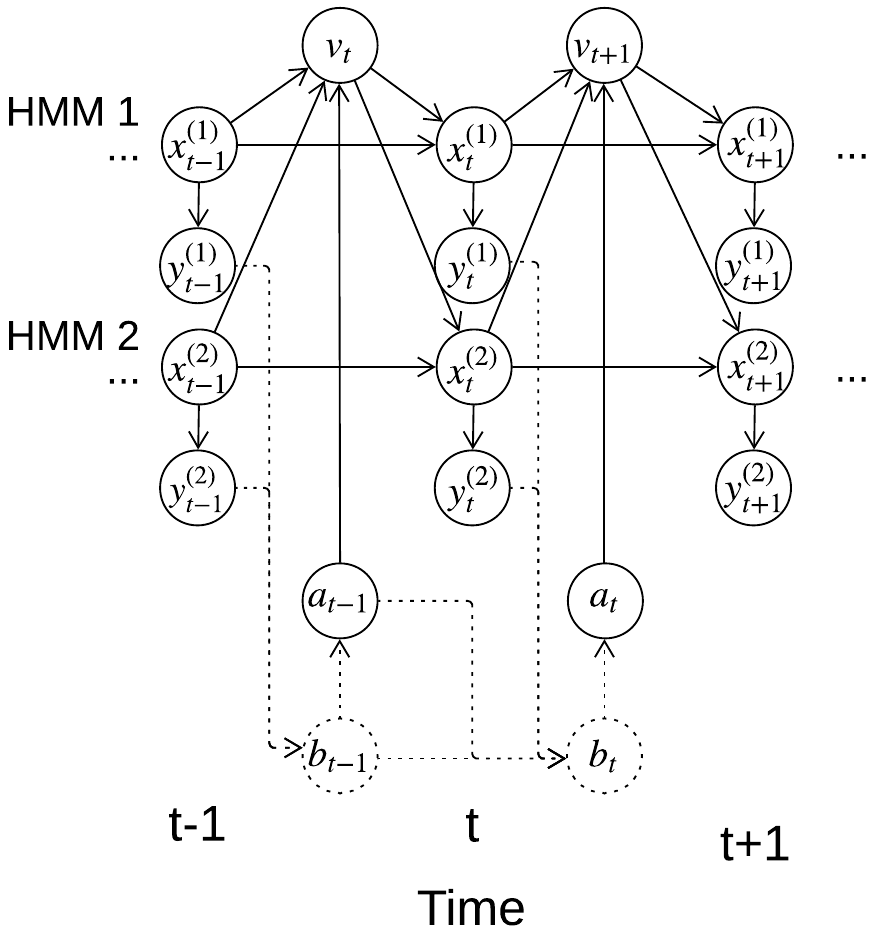} 
\label{fig:DEDP}}
\caption{\label{fig:Models} Graphical model representations of (\ref{fig:SKM}) a discrete event process and a (\ref{fig:DEDP}) a discrete 
event decision process.
A discrete event model captures complex 
system dynamics and specifies a decision-making problem compactly with a 
set of events.}
\end{figure}

A \textit{partially observable discrete-event decision process} (PODEDP) is a partially observable Markov decision process \cite{sutton2018reinforcement} where the dynamics are defined by a discrete-event model. Let $v_{t}$ be the \textit{event} happening at time \textit{t}. Let $x_{t}$ be the \textit{state} (populations of species or states of individuals), and $y_{t}$ the \textit{observation} about the state at time \textit{t}. Let $b_t = p(x_t|y_{t,t-1,\cdots}, a_{t,t-1}) = b_t(b_{t-1},y_t, a_{t-1})$ be the \textit{belief state} --- the probability distribution of the current system state estimated through observation-action history. Let $a_t$ be the \textit{control} (or \textit{action}) variables influencing the event-rate constants at time $t$, $c(a_t) = \left(c_1(a_t),\cdots,c_V(a_t)\right)$ where the action or its distribution is determined by the belief state $a_t = \pi(b_t;\theta)$ or $\pi = p(a_t|b_t)$. Further let $p(y_t|x_t)$ be the \textit{observation model}, $p(v_{t+1},x_{t+1}|x_t,a_t)$ the \textit{state transition model}, and $R(x_t,a_t)$ the \textit{immediate reward} at time $t$. The PODEDP problem (Eq. \ref{eq:podedp}) is to maximize the expected future reward of the discrete-time process defined by the probability measure $p(a_{0:t},v_{1:t},x_{0:t},y_{1:t})$ (Eq. \ref{eq:DTSKM}) by iteratively setting $a_{t}$ from a belief state $b_t$ that summarizes the observation-control history  $(y_{1:t-1},a_{0:t-1})$, where the indicator function $\delta(x_{t+1}\equiv x_{t}+\Delta_{v_{t+1}})$ is 1 if the current state is $x_{t+1}=x_{t}+\Delta_{v_{t+1}}$ and 0 otherwise, and $0<\gamma<1$ is a discount factor. The graphical model representations of a discrete-event process and a discrete-event decision process are given in Fig. \ref{fig:Models}.
\begin{align}
  & \text{arg 
max}_{a_{0:\infty}}\mathbf{E}_{x_{0:\infty},a_{0:\infty},v_{0:\infty},y_{0:\infty}}\left(\sum_{t=0}^{\infty}\gamma^t 
R(x_t,a_t) \right),\label{eq:podedp}\\
& \text{ where}\nonumber\\
  & p\left(a_{0:T},v_{1:T},x_{0:T},y_{1:T}\right) \label{eq:DTSKM}\\
  &\scriptstyle\hspace{.1in}= p(x_{0}) 
\prod_{t=0}^{T-1}p(a_{t}|b_{t})p(v_{t+1}|x_{t},a_{t})\delta(x_{t+1}\equiv 
x_{t}+\Delta_{v_{t+1}})p(y_{t+1}|x_{t+1}),\nonumber \\
  & p(v_{t+1}|x_{t},a_{t};\theta)=\begin{cases}
1-\sum_{k}\tau\cdot h_{k}(x_{t},a_{t}), & v_{t+1}=\emptyset\\
\tau\cdot h_{k}(x_{t},a_{t}), & v_{t+1}=k
\end{cases},\label{eq:DTSKMP}\\
&\scriptstyle h_{v}(x,c_{k})=c_{v}g_{v}(x)\text{ for 
}v=1,\cdots,V, \text{ and }h_{0}(x,c)=\sum\limits_{v=1}^{V}h_{v}(x,c_{v}). 
\label{eq:rate}
\end{align}
A Markov decision process (MDP) is a framework for modeling decision making in situations where outcomes are partly random and partly under the control of a decision-maker. It is used in many fields, including robotics, manufacturing, optimal control, game theory, and economics. In recent years, it has seen increasing applications in intelligent transportation systems, including traffic-signal control, autonomous driving, and traffic assignment.

\subsection{Modeling Traffic Dynamics with a Discrete-Event Model}\label{subsec:model_traffic_control}

In this subsection we define the traffic optimization problem in the framework of PODEDP (Eq. \ref{eq:podedp}), specifying the utility, states, events, observations, and control variables. We consider the movement of thousands of vehicles in a city-scale transportation network with $M$ locations and $V$ events, for which the typical length of a control episode is one day discretized into 1440 steps of one minute each. 

The \textit{state} $x_t=(x_t^{(1)},\cdots,x_t^{(M)},t)$ is the number of vehicles at the $M$ locations (road links and buildings) and the current time $t$. The \textit{observation} $y_{t}=(y_{t}^{(1)},\cdots,y_{t}^{(M)},t)$ is the number of probe vehicles at the $M$ locations and current time $t$, where the probe vehicles are randomly selected and constitute 10\% of the total vehicle population. The \textit{action} variables $a_t$ are the probability of choosing a downstream road link after completing the current road link, and the event rate coefficient of leaving or entering buildings. These action variables change the event rate coefficients to make road usage more efficient. The \textit{reward} function $R(x_t)=\sum_m\beta_{t,\text{perf}}^{(m)}x_t^{(m)}+\beta_{t,\text{trav}}^{(m)}x_t^{(m)}$ emulates the Charypar-Nagel scoring function \cite{horni2016multi} to reward performing the correct activities at locations and penalize traveling on roads, where $\beta_{t,\text{trav}}^{(m)}$ and $\beta_{t,\text{perf}}^{(m)}$ are the score coefficients.

Let $x_{\text{ttl}}$ be the total number of vehicles in the system and $y_{\text{ttl}}$ the total number of observed vehicles. The \textit{observation model} of observing $y_{t}^{(m_{j})}$ ``probe" vehicles at location $j$ conditioned on $x_{t}^{(m_{j})}$ vehicles in total is $p(y_{t}^{(m_{j})}|x_{t}^{(m_{j})})=C_{x_{t}^{(m_{j})}}^{y_{t}^{(m_{j})}}\cdot C_{x_{\text{ttl}}-x_{t}^{(m_{j})}}^{y_{\text{ttl}}-y_{t}^{(m_{j})}} / C_{x_{\text{ttl}}}^{y_{\text{ttl}}}$, where $C_a^b=a!/(b!\cdot (a-b)!)$ is $a$-choose-$b$. 

We implement the \textit{state transition model} $p(v_{t}, x_{t+1}\mid x_{t},a_{t})$ using discrete-event dynamics, where there is a single type of \textit{event} $p\cdot m_{1}{\scriptstyle \stackrel{c_{m_{1}m_{2}}}{\to}}p\cdot m_{2}$. A vehicle $p$ moving from one location $m_1$ to the next $m_2$ with event rate coefficient $c_{m_{1}m_{2}}$ changes the number of vehicles at the two locations to $x^{(m_1)}_t-1$ and $x^{(m_2)}_t+1$ respectively. The \textit{event rate coefficients} $c_{m_{1}m_{2}}$ (the probability of event per unit time as time goes to 0) of finishing the current road link is a function of the numbers of vehicles per lane per unit length, the maximum flow per lane, the speed limit, the length of the road link, and the road type (freeway, arterial road, or local road), implemented with a deep neural network and trained to best match the queue-network MATSim traffic dynamics \cite{horni2016multi}. In this way, we capture a variety of traffic behaviors involving traffic lights and traffic flow through a model-free approach.


Travel time modeled on a road link using an exponential random variable with matching average travel time is approximate. Nevertheless, such a model strikes a balance between capturing high-fidelity dynamics and being amiable to gradient-based machine learning algorithms. Because traffic-state estimation is driven by noisy observations, the inferred traffic state will be constrained by what the observations prescribe and thus will not stray far from the ground truth. This is different from pure simulation, where an approximate model can drive the system state to an unrealizable position. The optimal control from partial observations in this paper is a closed-loop control, such that any undesirable effect caused by model inaccuracy and randomness in the dynamics is corrected in the next time step. This approach is different from open-loop controls in classic transportation simulators for policy research.



\section{Tracking and Optimal Control}\label{sec:methodology}

In this section, we describe a particle filter algorithm that tracks traffic state (Alg. \ref{alg:PF}, Sec. \ref{subsec:traffic_prediction}) and implements optimal control from partial observations (Alg. \ref{alg:control}, Sec. \ref{subsec:traffic_control}).

\subsection{Tracking with Particle Filter}

\label{subsec:traffic_prediction}In this subsection, we derive a particle filter algorithm to track vehicles counts (belief state) at different road links and buildings  using the observed probe-vehicle counts at those locations in order to establish optimal control of transportation network dynamics. The usage of particle filtering permits to deal with noisy observations because it aggregates all information included in the past observations as a probability distribution of the current traffic state --- numbers of vehicles on the links and locations. We also derive particle smoothing to back trace the evolution of particles.

Let $y_{t}$ be a noisy observation of system state $x_{t}$ at time
$t$, and $a_{t}$ be the control. A particle filter (sequential Monte
Carlo method) approximates the posterior probability distribution
of a stochastic process through maintaining a collection of particles
$x_{t}^{k}$ and particle indices $i_{t}^{k}\in\{1,\cdots,K\}$ to
represent the likely system state  $x_{t}$, where $k=1,\cdots,K$
and $t=1,\cdots,T$. Inference with the particle filter involves tracking
the evolution of a stochastic process by alternating between particle
\emph{prediction} and \emph{updating}. In the \emph{prediction} step,
the particles at the next time step $t$ are sampled according to
the transition kernel $x_{t}^{k}\sim p(x_{t}\mid x_{t-1}^{i_{t-1}^{k}})$.
In the \emph{updating} step, the particle indices are resampled according
to the observation likelihood $i_{t}^{k}\mid x_{t}^{1},\dots,x_{t}^{K},y_{t}\sim\text{Categorical}\left(p(y_{t}\mid x_{t}^{1}),...,p(y_{t}\mid x_{t}^{K})\right)$.
With the resulting particles and particle indices, we use the empirical
probability $\frac{1}{K}\sum_{k}\delta_{x_{t}^{k}}(x_{t})$ to approximate
$p(x_{t}|y_{1},\dots,y_{t-1})$ and use $\frac{1}{K}\sum_{k}\delta_{x_{t}^{i_{t}^{k}}}(x_{t})$
to approximate $p(x_{t}|y_{1},\dots,y_{t})$, where $\delta$ is an
indicator function. 

To track a discrete event process initially at state $x_{0}$ at time
$t_{0}=0$, we initialize particle positions and indices as $x_{0}^{1},\cdots,x_{0}^{K}=x_{0}$
and $i_{0}^{1}=1,\cdots,i_{0}^{K}=K$, and alternate between the following
prediction step and updating step.

In the \emph{prediction} step, we sample particle positions $x_{t+1}^{1}\sim p(x_{t+1}|x_{t}^{i_{t}^{1}},a_{t}),\cdots,x_{t+1}^{K}\sim p(x_{t+1}|x_{t}^{i_{t}^{K}},a_{t})$
at time $t+1$ from the particles $x_{t}^{i_{t}^{1}},\cdots,x_{t}^{i_{t}^{K}}$
at time $t$. Specifically, we sample event $v_{t+1}^{k}$ according
to how likely it is that different events will occur conditioned on
system state $x_{t}^{i_{t}^{k}}$ for $k=1,\cdots,K$ and action $a_{t}$
(Eq. \ref{eq:pf-v}), and update $x_{t+1}^{k}=x_{t}^{k}+\Delta_{v_{t+1}^{k}}$
accordingly (Eq. \ref{eq:pf-x}). Because the resampled particles
are distributed according to $x_{t}^{i_{t}^{k}}\sim p(x_{t}|y_{1:t},a_{1:t-1})$,
the sampled particles are distributed according to $x_{t+1}^{k}\sim p(x_{t+1}|y_{1:t},a_{1:t})$.

The likelihood of particles $x_{t+1}^{k}$ are $p(y_{t+1}|x_{t+1}^{k})$
with respect to the observation $y_{t+1}$. To avoid particle degeneracy,
we perform a \textit{updating} step to eliminate particles with low
likelihood and duplicate particles with high likelihood (Eq.\ref{eq:pf-i}).
After the particle-updating step, all particles are distributed according
to $p(x_{t+1}|y_{1:t+1},a_{1:t})$, and all have the same likelihood.
\begin{align}
 & {\scriptstyle v_{t+1}^{k}|a_{t},x_{t}^{i_{t}^{k}}\sim\text{\small Cat}(1-\frac{h_{0}(x_{t}^{i_{t}^{k}},a_{t})}{\gamma},\frac{h_{1}(x_{t}^{i_{t}^{k}},a_{t})}{\gamma},\cdots,\frac{h_{V}(x_{t}^{i_{t}^{k}},a_{t})}{\gamma}),\label{eq:pf-v}}\\
 & {\scriptstyle x_{t+1}^{k}=x_{t}^{i_{t}^{k}}+\Delta_{v_{t+1}^{k}},\label{eq:pf-x}}\\
 & i_{t+1}^{k}|x_{t+1}^{1:K},y_{t+1}\!\sim\!\text{Cat}(p(y_{t+1}|x_{t+1}^{1}),\!\cdots\!,p(y_{t+1}|x_{t+1}^{K})).\label{eq:pf-i}
\end{align}

To derive a particle trajectory from the posterior distribution of a stochastic kinetic process with respect to observations, we trace back the events that lead to the particles $x_{T}^{i_{T}^{k}}$ for $k=1,\cdots,N$:
\begin{align}
  & x_{0},a_{0},v_{1}^{j_{1}^{k}},x_{1}^{j_{1}^{k}},a_{1},\cdots,v_{T}^{j_{T}^{k}},x_{T}^{j_{T}^{k}},a_{T}, 
\label{eq:pf-path}\\
&\hspace{.1in}\text{where } 
j_{T}^{k}=i_{T}^{k},j_{T-1}^{k}=i_{T-1}^{j_{T}^{k}},j_{T-2}^{k}=i_{T-2}^{j_{T-1}^{k}},\cdots,j_{1}^{k}=i_{1}^{j_{2}^{k}}. \nonumber
\end{align}

The particles $x_{t}^{i_{t}^{1}},\cdots,x_{t}^{i_{t}^{K}}$ form an approximation of the forward probability $p(x_{t}|y_{1,\cdots,t})$ and likelihood $p(y_{1,\cdots,T})$. The ancestral lines of the particles 
$x_{T}^{i_{T}^{k}}$, where $k=1,\cdots,K$, form an approximation of the posterior distribution of the stochastic process conditioned on observations, where $\delta$ is an indicator function:
\begin{align}
  & \hat{p}(x_{t}|y_{1:t})=\frac{1}{K}\sum\limits 
_{k}\delta(x_{t}^{i_{t}^{k}}\equiv 
x_{t})\stackrel{K\to\infty}{\longrightarrow}p(x_{t}|y_{1:t}),\label{eq:belief}\\
  & 
\hat{p}(y_{1:T})=\prod_{t}\hat{p}(y_{t}|y_{1:t-1})=\scriptstyle\prod\limits_{t}\frac{1}{K}\sum\limits_{k}p(y_{t}|x_{t}^{k})\stackrel{K\to\infty}{\longrightarrow}p(y_{1:T}),\label{eq:loglik}\\
  & \hat{p}(x_{1:T}|y_{1:T})=\scriptstyle \frac{1}{K}\sum\limits 
_{k}\delta((x_{1:T}^{j_{1:T}^{k}})\equiv 
(x_{1:T}))\stackrel{K\to\infty}{\longrightarrow}p(x_{1:T}|y_{1:T}).
\end{align}

Overall, we develop a particle-based algorithm to update the belief state and calibrate the parameters of a discrete event decision process (Algorithm \ref{alg:PF}).

\begin{algorithm}[t]

\textbf{Input}: Observations $y_{1},\cdots,y_{T}$ and control inputs 
$a_1,\cdots,a_T$
of a discrete event decision process (Eq. \ref{eq:DTSKM}).

\textbf{Output}: Belief state $b_t\approx\hat p(x_t|y_{1:t})$ (Eq. \ref{eq:belief}) where particles 
$(v_{t}^{i_{t}^{k}},x_{t}^{i_{t}^{k}})_{t=1:T}^{k=1:K}$ are sampled
from particle filter, and particle trajectories 
$(v_{t}^{j_{t}^{k}},x_{t}^{j_{t}^{k}})_{t=1:T}^{k=1:K}$
are sampled from particle smoother.
\vspace*{.5em}

\textbf{Procedure}:
\begin{itemize}
\item Initialize $x_{0}^{1}=\cdots=x_{0}^{N}=x_{0}$, 
$i_{0}^{1}=1,\cdots,i_{0}^{K}=K$.
\item (Filtering) For $t=1,\cdots,T$ and $k=1,\cdots,K$, sample $v_{t}^{k}$
and $i_{t}^{k}$ according to Eq. \ref{eq:pf-v}, \ref{eq:pf-x} and
\ref{eq:pf-i}.
\item (Smoothing) Back-track particle trajectory from $x_{T}^{i_{T}^{k}}$
according to Eq. \ref{eq:pf-path}, for $k=1,\cdots,K$.
\end{itemize}

\textbf{Calibration}: 
Maximize log likelihood $\log\hat p(y_{1:T})$ (Eq. \ref{eq:loglik}) with gradient ascent.
\caption{\label{alg:PF}Particle filtering, smoothing, and parameter 
learning for discrete event decision process}
\end{algorithm}

\subsection{Optimal Control with Particle Filter}

\label{subsec:traffic_control}In this subsection, we derive a particle-based algorithm to identify the optimal control of a complex system from our estimation of the current system state (belief state), using the equivalence between the state-value function of a Markov decision process and the probability of receiving the reward from a mixture of finite-time Markov decision 
processes \cite{toussaint2006POMDP}. This equivalence enables the translation of the policy-evaluation and policy-improvement steps in a policy iteration algorithm into the expectation and maximization steps in an expectation maximization (EM) algorithm, and the application of a large variety of approximate inference algorithms for dynamic Bayesian networks to solve 
intractable optimal control problems. In particular, it is based on the following derivation:
\begin{align}
  & 
\mathbf{E}\sum_{t=0}^{\infty}\gamma_{t}R(a_t,x_t)=\sum_{t=0}^{\infty}\gamma_{t}\mathbf{E}\left(R(a_t,x_t)\right) \label{eq:V}\\
& \hspace{1in}\propto\sum _{T=0}^{\infty}p(T)\sum 
_{t=0}^{T}p(t)\mathbf{E}(p(R=1|a_t,x_t)), \nonumber\\
  &  \text{where }\gamma_{t}\!\propto\!\!\sum 
_{T=0}^{\infty}p(T)p(t)\delta_{t\le T}, p(R=1|a_t,x_t)\!\propto\! R(a_t,x_t).\nonumber
\end{align}

Eq. \ref{eq:V} connects the expected future reward of a Markov decision process and the probability of receiving a binary reward in a mixture of finite time Markov decision processes $(T, t,\xi_{t},R)$. This finite time Markov decision process executes the same plan as the original Markov decision process up to a terminal time $t$, it generates a state-action trajectory $\xi_{t}=x_{0},a_{0},\cdots,x_{t},a_{t}$, and it receives a binary reward with probability $p(R=1|x_t,a_t)\propto R(x_t,a_t)$. In Eq. \ref{eq:V}, $\gamma_{t}$ is a discount factor. Corresponding to the expected discounted 
cumulative future reward with $\gamma_{t}=\gamma^{t}$, we select $p(T)=(1-\delta)\delta^{t}$ and $p(t)=(1-\frac{\gamma}{\delta})(\frac{\gamma}{\delta})^{t}$. Corresponding to the expected finite-horizon future reward with 
$\gamma_{t}=\delta(t\le H)$, we select $p(T)=\delta(H\equiv T)$ and $p(t)=1/(H+1)$, where indicator function $\delta(T\equiv H)=1$ when $T=H$ and $0$ otherwise, and $\delta(t\le H)=1$ when $t\le H$ and $0$ otherwise.

To identify optimal control with the EM algorithm in a discrete event decision process, we maximize the expected log likelihood $\mathbf{E}_{T,t,\xi_t}\log p(T,t,\xi_t,R=1;\theta)$ by alternately identifying the typical state-action sequences generated by a policy that leads to reward ($p(T,t,\xi_t|R=1;\theta)$) in the expectation step (E-step) and tuning the control parameters of the policy ($\theta$) so that these typical sequences lead to reward with higher probabilities in the maximization step (M-step). The EM algorithm is an iterative algorithm that searches for the parameters to maximize the expected log likelihood over the posterior probability distribution of the latent variables conditional on the 
observations. Here, the likelihood is proportional to the value function, the latent variables are a sequence of states and actions, and the observations are of whether a reward is received.

In E-step, we use importance sampling to approximate the proxy of future expected reward $p(R=1)$ and the posterior probability $p(x_\tau,a_\tau \mid R=1,\tau\le t)$ induced in Eq. \ref{eq:V}. Specifically, we sample $T^{k}\sim p(T)$ and $\xi^{k}\sim p(\xi_{T}|T^{k})$ for $k=1,\cdots,K$, we approximate the prior distribution $p(T,\xi_{T})$ 
with sample distribution $\hat{p}(T,\xi_{T})=\frac{1}{K}\sum_{k=1}^{K}\delta(T^{k},\xi^{k})\equiv (T,\xi_{T}))$, and use importance weight $\sum_{t=0}^T p(t)p(R=1|x_t^k,a_t^k)$ to approximate the posterior distribution, where $\delta$ is an indicator function.
\begin{align}
& T^k,a_{0:T^k}^k,x_{0:T^k}^k,v_{1:T^k}^k \sim  \label{eq:E}\\
&\hspace{.0in} P(T^k) b(x_0) 
\prod_{t=0}^{T^k} p(a_t|x_t;\theta) 
p(v_t|a_t,x_t)\delta(x_{t+1}\equiv x_t+\Delta_{v_{t+1}}) \nonumber\\
& \scriptstyle\mathbf{E}\sum\limits_{t=0}^{\infty} \gamma_{t}R(a_t,x_t) \propto  
p(R=1) \approx \frac{1}{K}\sum\limits_{k,t} 
p(t)p(R=1|x_{t}^{k},a_{t}^{k}) \label{eq:V-pf} \\
  & \hat{p}(x_{\tau}=x,a_{\tau}=a|R=1,\tau\le t)\label{eq:XA}\\
  &\hspace{.1in}= \frac{\sum\nolimits 
_{k=1}^{K}\delta(x\equiv x_{\tau}^{k}) \delta(a\equiv a_{\tau}^{k}) 
\sum\nolimits _{t=\tau}^{T}p(t)p(R=1|x_{t}^{k},a_{t}^{k})}{\sum\nolimits 
_{k=1}^{K}\sum\nolimits 
_{t=\tau}^{T}p(t)p(R=1|x_{t}^{k},a_{t}^{k})}.\nonumber
\end{align}

The posterior probability (Eq. \ref{eq:XA}) is the fraction of expected future discounted reward received from $x_{\tau}^{k}=x,a_{\tau}^{k}=a$ over the total expected future discounted reward received after $\tau$, 
averaged over sample paths $\{(T^k,\xi^k):k\}$.

In M-step, we iteratively maximize the expected log likelihood of receiving a reward. The optimal control $\theta$ is consequently set such that actions $a$ appears in proportion to the future rewards.
\begin{align}
  & \mathbf{E}_{\theta^{\text{old}}}\log p(\xi_T,T, 
R=1;\theta)=\cdots+\mathbf{E}_{\pi^{\text{old}}}\log 
p(a_{t},x_{t}|R=1;\theta),\nonumber \\
  & \Rightarrow p(a|x;\theta)= \frac{\scriptstyle\sum\limits 
_{k,\tau}\delta(x_{\tau}\equiv x)\delta(a_{\tau}\equiv a)\sum\limits 
_{t=\tau}^{T^k}p(t)p(R=1|x_{t}^{k},a_{t}^{k})}{\scriptstyle\sum\limits 
_{k,\tau}\delta(x_{\tau}\equiv x)\cdot\sum\limits 
_{t=\tau}^{T^k}p(t)p(R=1|x_{t}^{k},a_{t}^{k})}.\label{eq:M}
\end{align}

To summarize, we develop an algorithm \ref{alg:control} to control a complex system from a discrete event model and noisy observations.

\begin{algorithm}[th]
\textbf{Input:} Belief state (Eq. \ref{eq:belief}) of discrete event process (Eq. \ref{eq:DTSKM}) at time 0 as particles, $\{{x}_{0}^{k}:k=1\text{:}K\}$.
Initial policy $\theta$ at time 0.

\textbf{Output:} Optimal action $a_t$ according to Eq. \ref{eq:M}.

\textbf{Calibration:} Improve policy $\theta$ through policy iteration. 

\begin{itemize}
\item E step. For $k=1\text{:}K$: sample $T^k,\xi^k$ according to Eq. \ref{eq:E}
\item M step. Upate $\theta$ according to Eq. \ref{eq:M}. 
\end{itemize}
\caption{\label{alg:control}Optimal control from belief state with Particle Filter}
\end{algorithm}

\section{Tracking and Planning in City-Scale Transportation Networks}\label{experiments}

In this section, we benchmark our framework with other state-of-the-art algorithms on tracking and optimizing the travel plans in a city-scale transportation network from noisy observations of network dynamics.

\subsection{Data Description}

We evaluate the performance of our framework on five datasets of human mobility: (1) SynthTown, (2) Berlin, (3) Santiago de Chile, (4) Dakar, and (5) NYC Taxicab. We use as varied data as possible to demonstrate that our proposed framework is useful in more than a narrow range of cases and is fair. The different cases indeed show different levels of uncertainty in traffic state estimation, different travel times, and different on-time arrival rates. Nevertheless, the proposed traffic tracking and control algorithms outperforms the state of the art.

\begin{figure}
\centering\includegraphics[width=0.90\columnwidth,height=.55\columnwidth]{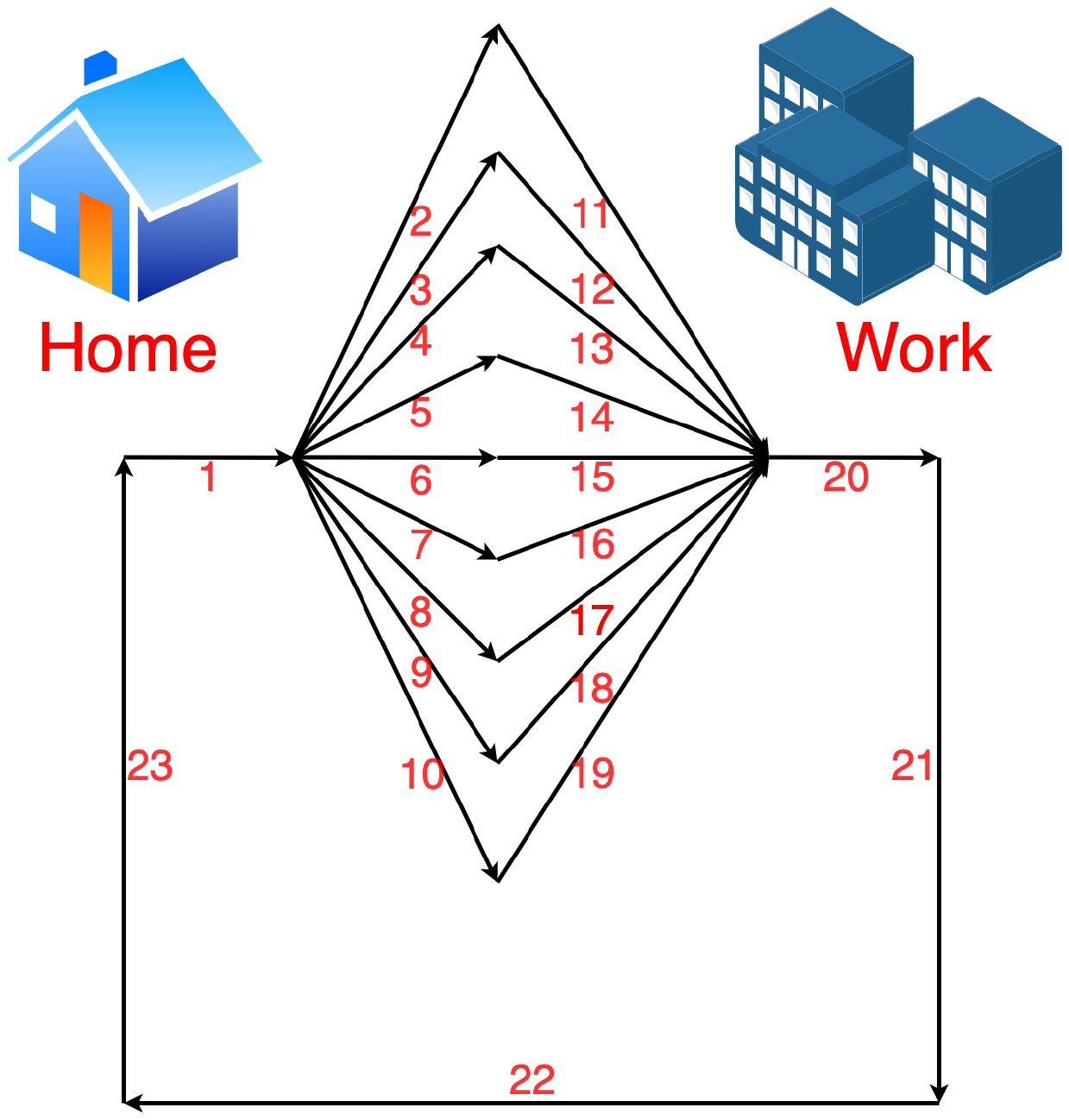}
\caption{\label{fig:SynthTown} SynthTown road network, which contains 2 facilities, Home and Work, and 23 road links labeled from 1 to 23.}\vspace{-.2in}
\end{figure}

The SynthTown dataset is comprised of a synthesized network of one home location, one work location, and 23 single-direction road links (Fig. \ref{fig:SynthTown}) to characterize the trips of 2000 synthesized inhabitants going to work in the morning and returning home in the afternoon \cite{horni2016multi}. The prediction problem is to estimate the vehicle counts at home, at work, and at links 1-23 in the present time, 10 minutes later, and 60 minutes later from observations of the 200 ``probe'' inhabitants collected at link 1 and link 20. The control problem is to maximize a proxy of the Charypar-Nagel scoring function \cite{horni2016multi} from setting control variables according to these observations. We use this dataset to show the details of tracking and  control results.  

The Berlin dataset is comprised of a network of 11 thousand nodes and 24 thousand single-direction car-only links derived from OpenStreetMap; and the trips of 9 thousand synthesized vehicles representing the travel behaviors of three million inhabitants \cite{ziemke2015integrating}. To make the problem small enough that algorithms with bigger time-complexity can run and have performances compared with our algorithm, we aggregate the 24 thousand road links into 1539 clusters with a walk-trap algorithm \cite{pons2006computing}.
The synthesized daily trips have been validated based on extensive, regularly-conducted travel surveys and constitute a quality representation of road transport demand. This data set is the result of a generalizable approach to synthesize individual-level behaviorally-sound trip diaries from easily accessible input data, since collecting the trip diaries of real-world people is plagued with privacy issues.

The Santiago de Chile dataset is comprised of a network of 23 thousand nodes and 38 thousand single-direction car-only links derived from OpenStreetMap; and the trips of 665 thousand synthesized vehicles representing the travel behaviors of six million inhabitants in car, walking and public transportation modals \cite{kickhofer2016creating}. The daily trips in the Santiago de Chile dataset were initialized from cloning the sequences of activities (starting time and duration of home, work, school, shopping, leisure, visit and health) and travel mode of 60 thousand individuals (from 18 thousand households) from publicly-accessible travel diaries, and modified through physical simulation and a co-evolutionary algorithm (with MATSim) to maximize the overall utility of the system. The resulting daily trips are compatible with travel modals' distributions and observed traffic counts at count stations. This data set represents the case where we can get travel diaries with fine temporal and spatial resolution for a significant and representative fraction of a population from publicly-accessible travel diaries.

The Dakar dataset is comprised of a network of 8 thousand single-direction road links derived from OpenStreetMap and 12 thousand real-world vehicle trips derived from the ``Data for Development (D4D)'' data sets based on the Call Detail Records (CDR) of over 9 million Sonatel customers in Senegal (out of 15 million total population) through year 2013 \cite{de2014d4d}.
From data set 2, we identify the home and work/school locations of each user as randomly picked locations from the most appeared sites during 7am - 7pm and 7pm - 7am respectively. Then, we sample an activity-trip sequence for each user to match her/his sequence of mobility records (in data set 2) from a Markov chain model describing how s/he performed various activities (home, work, school, shopping, etc). This data set represents the case where we can get travel diaries with fine temporal and spatial resolution for a significant and representative fraction of a population through mobile phones. 

The NYC TaxiCab dataset\footnote{http://nyc.gov/tlcopendata} is comprised of a network of 7 thousand nodes and 11 thousand single-direction road links derived from OpenStreetMap and an average of 1 million daily trips of taxicabs and for-hire vehicles (including Uber, Lyft, Via and Juno) throughout 2018. Each trip record contains pick-up and drop-off zones among the 236 zones in New York City, and pick-up and drop-off data and time, among other information. The trip records are made publicly accessible by the New York City Taxi and Limousine Commission (an agency responsible for licensing and regulating New York City's taxi cabs, for-hire vehicles, commuter vans, and paratransit vehicles). Together with many other open data sets through the City's Open Data portal\footnote{https://opendata.cityofnewyork.us/}, TLC's trip data has a big impact in making the city street smart. Here, we use the data to predict the behavior of all taxicabs and for-hire vehicles from observing a small fraction of them.

\subsection{Tracking Transportation Dynamics}

\textbf{Benchmark algorithms}: We firstly benchmark our framework --- stochastic kinetic model with particle filter (PFSKM) --- against a Deep Neural Network (DNN) \cite{lv2015traffic,polson2017deep}, a Recurrent Neural Network (RNN) \cite{zhao2017lstm}, and an extended Kalman filter (EKF) \cite{wang2006renaissance,van2012localized} in the task of continuously tracking the current and future traffic conditions. DNN represents the power of a general-purpose non-parametric model. We build a five layer Deep Neural Network (DNN): (i) an input layer accepting the observation history of probe vehicles at selected locations, (ii) three hidden layers, and (iii) one output layer generating the inferred distribution of all vehicles at all locations. RNN exploits the temporal structure that recursively takes the inferred result from the previous cell as well as the current observations as input, and output the estimated vehicle distribution. Both DNN and RNN are trained with 30 days of synthesized mobility data from MATSim until obtaining optimum performance. The EKF, instead, assumes a Gaussian distribution between the time-indexed latent states, and we implement a standard EKF procedure that alternates between predicting and updating steps. We trained the DNN and RNN models with stochastic gradient descent, and trained the EKF model with expectation maximization.

\textbf{Evaluation metric}: We use two metrics to evaluate the performance of our model: coefficient of determination ($R^{2}$) and mean squared error (MSE). We use $R^{2}$ to evaluate the goodness of fit between a time series of the estimated vehicle counts at a location and the ground truth. Let $f_{t}$ be the estimated vehicle count at time $t$, $y_{t}$ the ground truth and $\bar{y}$ the average of $y_{t}$. We define $R^{2}=1-\sum_{t}(f_{t}-y_{t})^{2}/\sum_{t}(y_{t}-\bar{y}_{t})^{2}$.
A higher $R^{2}$ indicates a better fit between the estimated time series and the ground truth, with $R^{2}=1$ indicating a perfect fit and $R^{2}<0$ a fit worse than using the average. We use MSE to measure the average squared error difference between the estimated vehicle counts at all locations at a time $t$ and the ground truth. A lower MSE represents a more precise prediction. Let $f^{(i)}$ be the estimated vehicle count at location $i$ and $y^{(i)}$ the ground truth. We define $\operatorname{MSE}={\frac{1}{n}}\sum_{{i=1}}^{n}({y^{(i)}}-f^{(i)})^{2}$. 

\begin{figure}
\subfloat[Ground truth]{\includegraphics[width=0.48\linewidth]{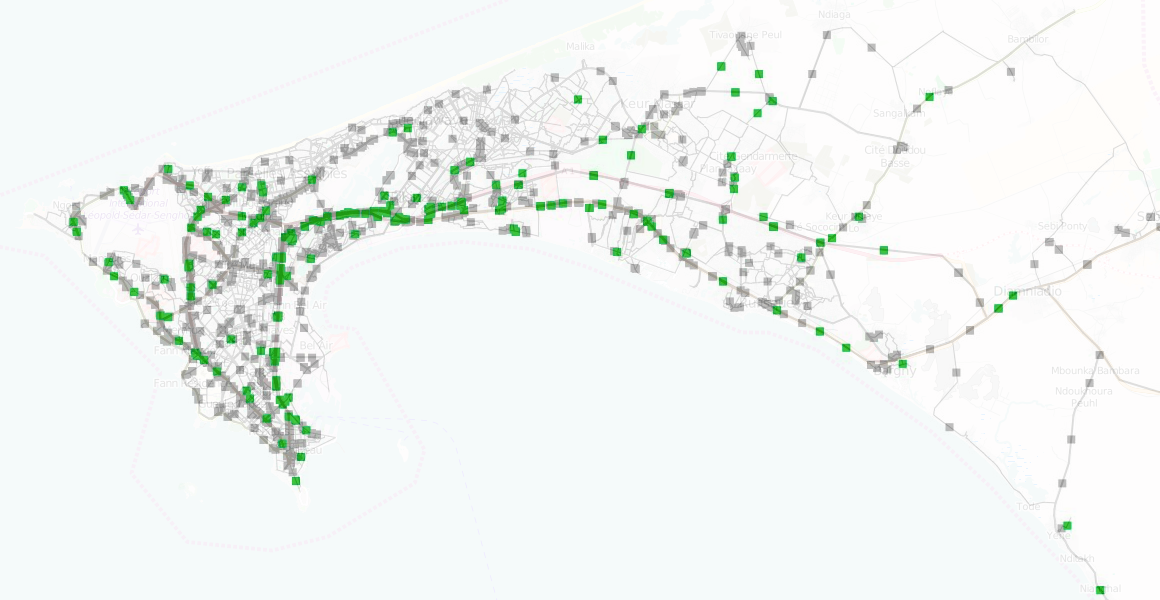}}
\subfloat[Particle filter]{\includegraphics[width=0.48\linewidth]{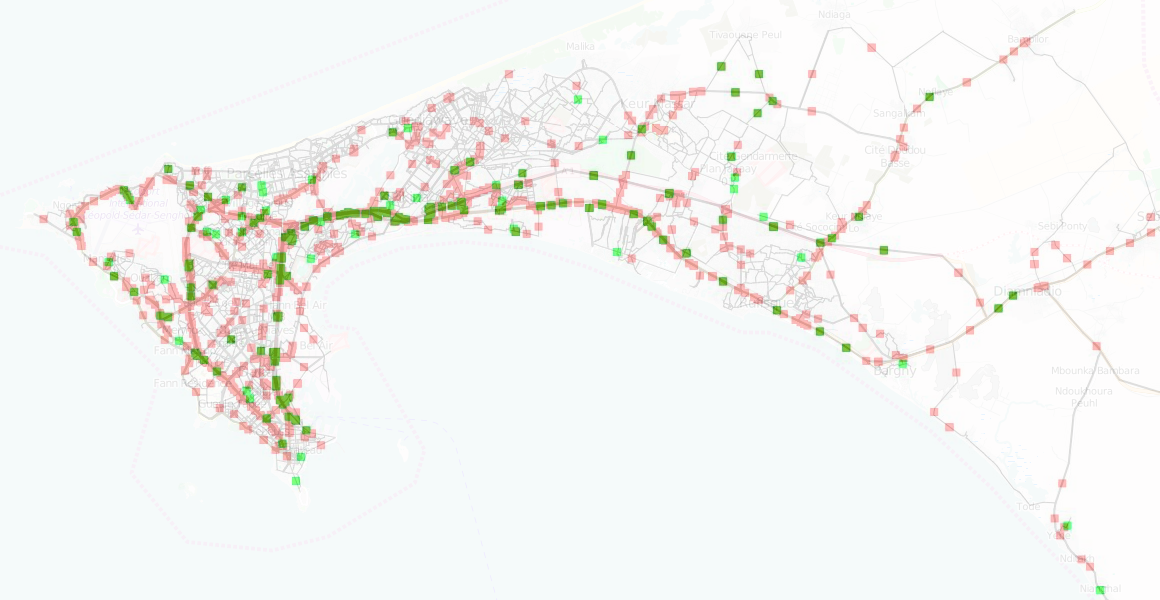}}
\caption{\label{fig:visualization}Comparing estimated and ground
truth traffic densities through two snapshots taken at the same time (black dots: vehicles in the ground truth; red dots: vehicles from a particle trajectory; green dots: probe vehicles).}\vspace{-.2in}
\end{figure}


\textbf{Results' visualization}: One benefit of using a discrete-event model with particle filter is that we can qualitatively visualize how vehicles move in a city-scale road network in accordance with ``probe'' vehicle locations, and how traffic policies change vehicle behavior.
Fig. \ref{fig:visualization} compares the distributions of vehicles in the ground truth and in a particle trajectory (Eq. \ref{eq:pf-path}) using as an example the Dakar data set through two snapshots taken at the same time. It can be observed from the figures that the vehicle density in ground truth and estimation agree with each other, and
both are proportional to the density of ``probe'' vehicles.

\begin{figure*}
\subfloat[SynthTown  MSE]{\includegraphics[width=0.33\linewidth]{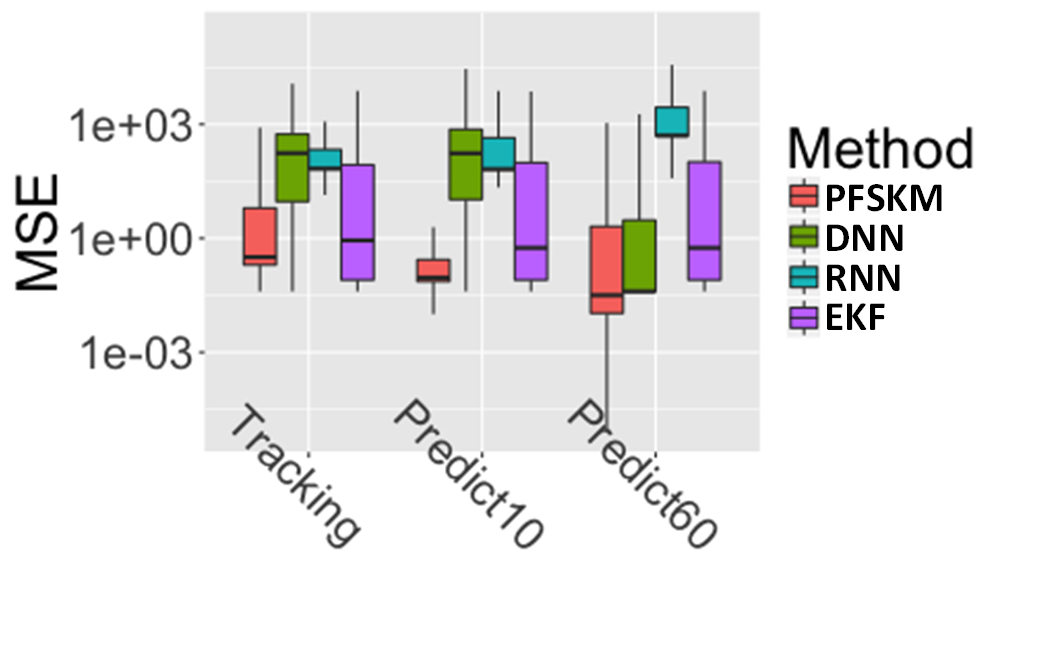}}
\subfloat[Berlin MSE]{\includegraphics[width=0.33\linewidth]{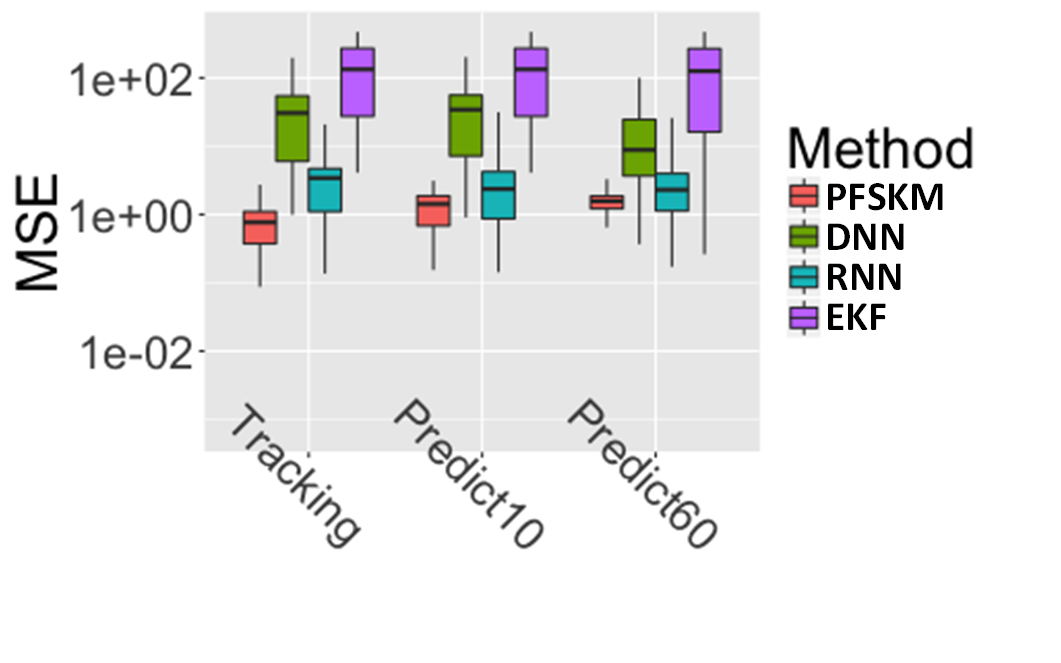}}
\subfloat[Santiago, Darkar, NYC  MSE with PFSKM]{\includegraphics[width=0.29\linewidth]{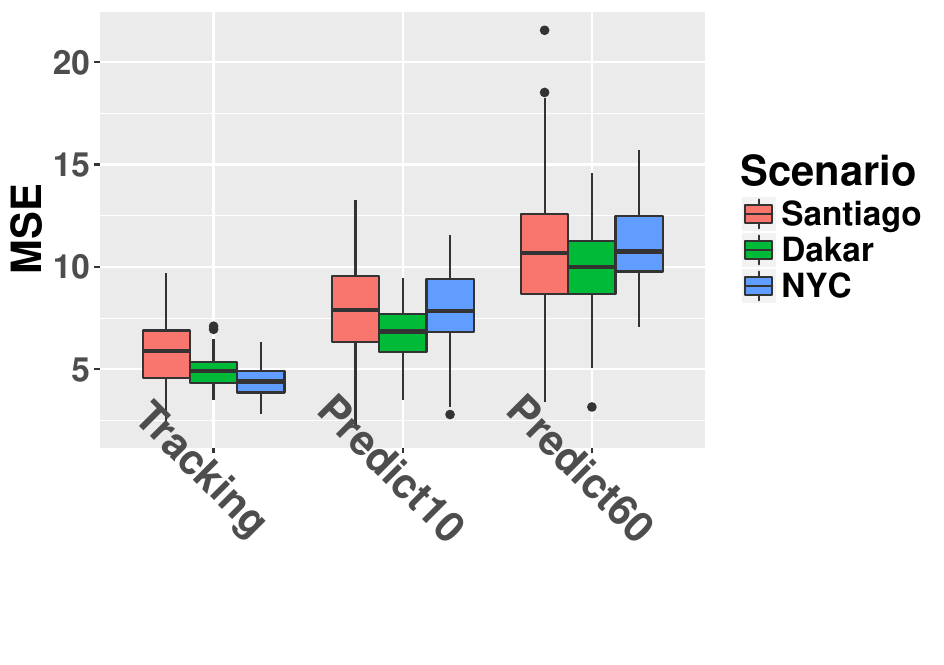}}\\
\subfloat[SynthTown  $R^{2}$]{\includegraphics[width=0.33\linewidth]{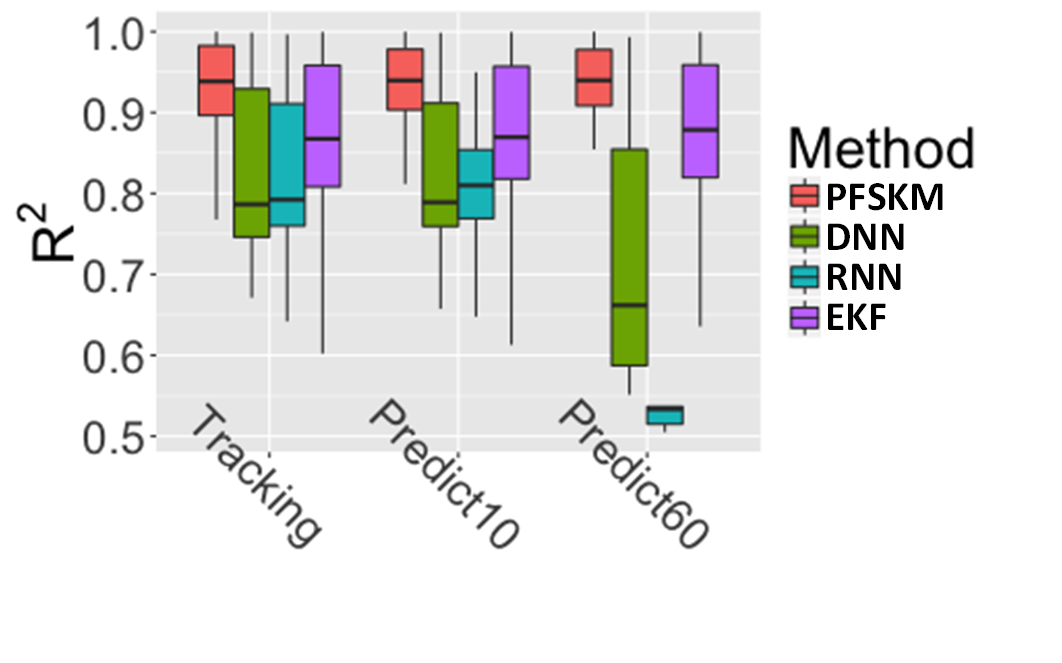}}
\subfloat[Berlin $R^{2}$]{\includegraphics[width=0.33\linewidth]{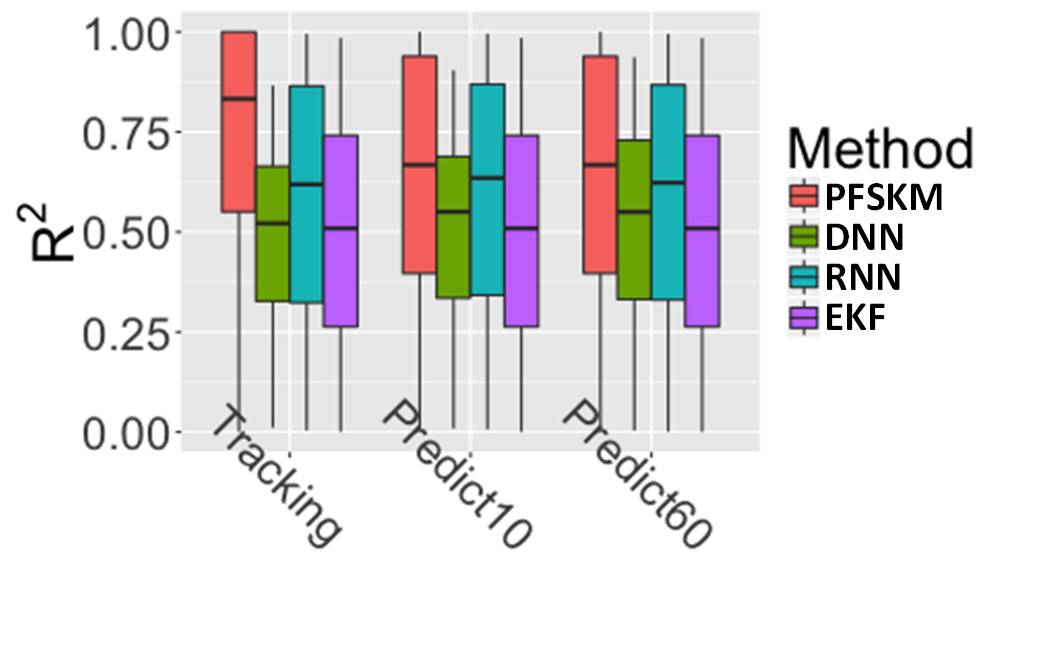}}
\subfloat[Santiago, Darkar, NYC  $R^{2}$ with PFSKM]{\includegraphics[width=0.30\linewidth]{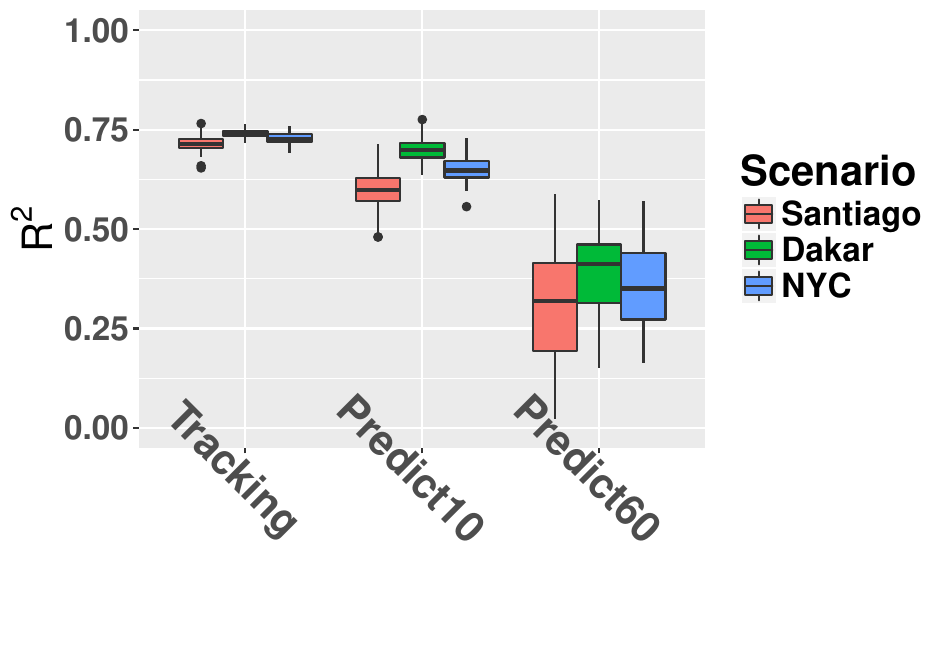}} \caption{\label{fig:performance_comparison}Performance of PFSKM, DNN, RNN
and EKF on SynthTown, Berlin, Dakar, Santiago de Chile and NYC Taxicab datasets using MSE (top, lower MSE indicates better performance) and $R^{2}$ (bottom, higher $R^{2}$ indicates better performance).}\vspace{-.3in}
\end{figure*}

\textbf{Evaluation results}: Figure \ref{fig:performance_comparison} summarizes the MSE and $R^{2}$ performance statistics of the four models for a vehicle-tracking task --- estimating the numbers of vehicles up to now, with short-term (10 minutes) and long-term prediction (1 hour) on all the datasets. The Dakar dataset is simply too large for DNN, RNN, and EKF, which demonstrates the superior scalability of PFSKM. PFSKM has the lowest MSE across different times of day, followed in order by DNN, EKF, and RNN (top row, lower is better). PFSKM also has the highest $R^{2}$ across different locations, followed by DNN, EKF, and RNN (bottom row, higher is better). PFSKM outperforms RNN and DNN because it can explicitly leverage problem-specific structures such as road topology. PFSKM outperforms EKF because it can work with arbitrary probability distributions, and sometimes a Gaussian assumption is not a good approximation for real-world applications. This comparison also points to new developments in neural network architectures that are either regularized by event-based structures of a complex system or can learn such structures explicitly.

\begin{figure}
\centering\includegraphics[width=0.9\columnwidth, height=0.85\columnwidth]{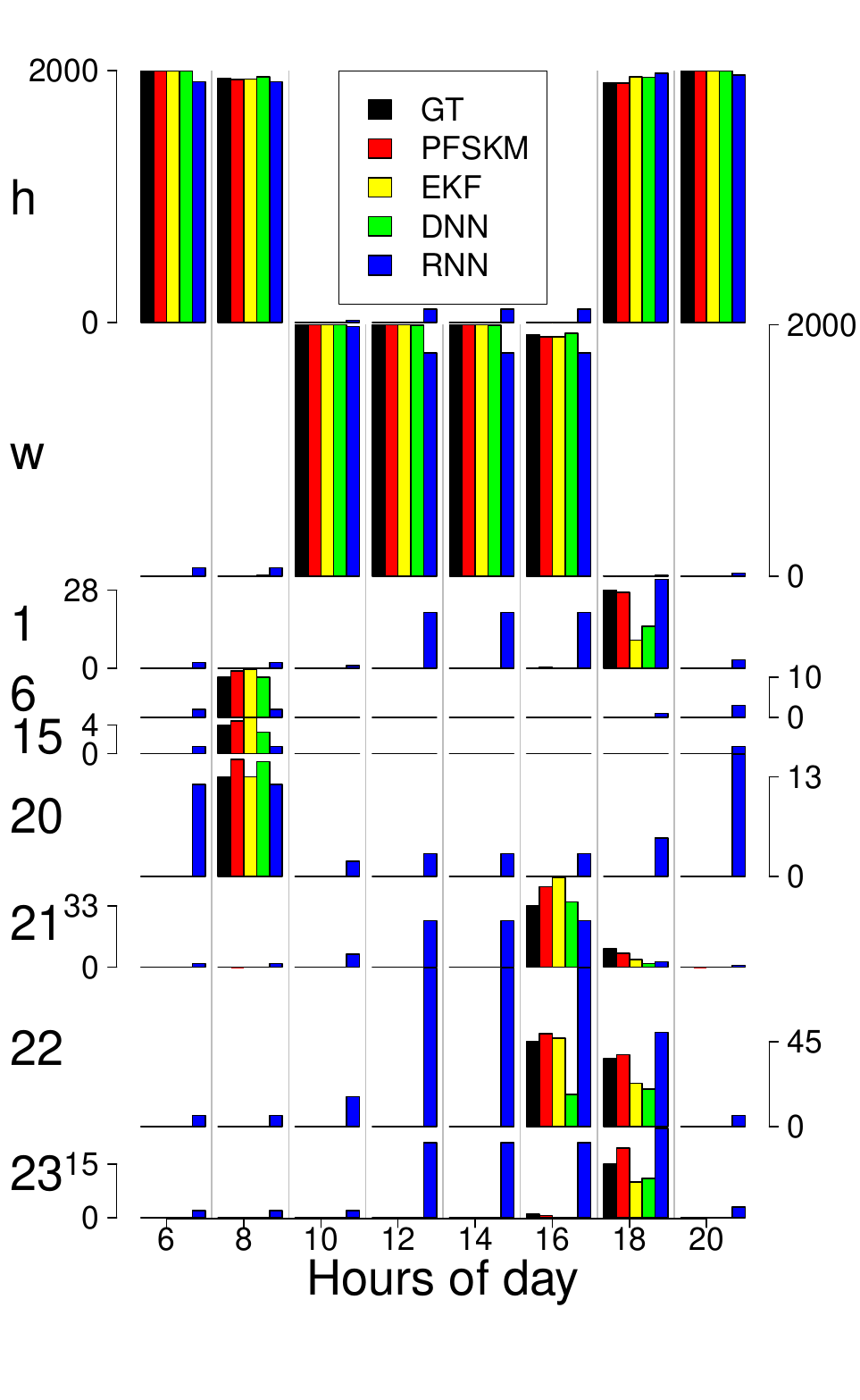}
\caption{\label{fig:SynthTownPrediction} Predicted vehicle counts at home, work, and different road segments of SynthTown (y-axis, with locations marked on the left) at different hours of a day (x-axis) from the observations of probe vehicles (10\% of the total) at links 1 and 20 only. Algorithms with a good performance should lead to a prediction closer to the ground truth (GT).}\vspace{-.27in}
\end{figure}

\textbf{Comparing detailed predictions using SynthTown data}: Fig. \ref{fig:SynthTownPrediction} shows how PFSKM, DNN, RNN, and EKF predict the number of vehicles at different locations of SynthTown one hour ahead of time throughout the day from observations of probe vehicles (10\% of the total) at links 1 and 20 only. The x-axis indicates the hour of the day, the y-axis shows the number of vehicles at different locations (home, work, and road segments marked on the left), and the ground truth (GT) serves as the frame of reference. 

All four algorithms perform well, indicating that they all learn the structure in the dynamics. In fact, there is little uncertainty about the traffic dynamics at SynthTown if the number of vehicles on links 1 and 20 can be monitored, although with noise. RNN underperforms the other three algorithms because learning the structure of a dynamic system requires a huge training dataset. PFSKM estimation agrees with GT, and is closer than DNN and RNN estimations. This is because PFSKM explicitly leverages problem-specific structures such as road topology, while DNN and RNN must learn them implicitly and gradually. PFSKM is more accurate than EKF estimation because PFSKM can work with arbitrary probability distributions while EKF assumes Gaussianity. EKF and DNN agree well with GT at busy locations (home and work) but less well at locations with few people, which demonstrates that PFSKM better adopts dynamic changes.

\subsection{Optimal Control in Transportation Dynamics}

\begin{figure}
\includegraphics[width=.9\linewidth]{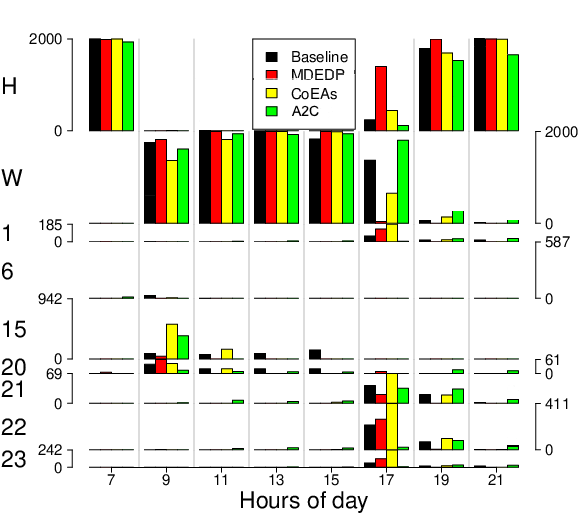}
\caption{\label{fig:SynthTownDynamics} Vehicle counts at home, work, and various road segments of SynthTown (y-axis, with locations marked on the left) at different hours of a day (x-axis) after executing different planning and control algorithms from the observations of probe vehicles (10\% of the total) at links 1 and 20 only. Algorithms with a good performance should lead to lower trip time and higher on-time arrival ratio.}\vspace{-.25in}
\end{figure}


\textbf{Benchmark algorithms}: In the previous section we have demonstrated the tracking capability of our framework with particle filter. Now, we evaluate our framework for optimizing traffic against (i) a within-day re-planning baseline algorithm \cite{illenberger2007enhancing}, (ii) a co-evolutionary algorithm implementing open-loop control \cite{horni2016multi}, and (iii) an advantage actor-critic algorithm \cite{konda2000actor} implementing closed-loop control. The baseline algorithm (Baseline) optimizes agents' expected future rewards by considering the current traffic situation but not the plans of other agents. The co-evolutionary algorithm (CoEA) is the state-of-the-art algorithm for generating the equilibrium of daily activities and trips in transportation theory \cite{horni2016multi}. In CoEA, agents independently explore and exploit their plans through a genetic operator, then jointly execute and evaluate their plans in a simulator, and finally repeat this process until an equilibrium is reached \cite{popovici2012coevolutionary}. The advantage actor-critic algorithm (A2C) uses a ``critic'' to estimate the traffic situation in terms of an action-value function and an ``actor'' which suggests optimal departure times and route choices in the direction suggested by the critic. We further use advantage to lower the variance.

\textbf{Evaluation metric}: We use three metrics to evaluate the different planning algorithms. The first is \textit{average trip time} in minutes of all vehicles driving from home to work: a lower average trip time means better traffic. The second is \textit{on-time arrival ratio}, which measures the percentage of people arriving to work on time. Finally, we use \textit{expected reward} per vehicle per hour, where higher expected rewards mean better individual plans and a more efficient transportation network.

\textbf{Comparing detailed behaviors on SynthTown data}: Figure \ref{fig:SynthTownDynamics} shows the vehicle counts at the different locations of SynthTown throughout the day after executing different planning algorithms from the observations
of probe vehicles (10\% of the total) at links 1 and 20 only. The x-axis indicates the hours of the day, the y-axis shows the numbers of vehicles at different locations (home, work, and road segments marked on the left), and the baseline
(Baseline) serves as the frame of reference. Note that vehicles applying the policy from our framework best satisfy the requirements of all individuals. Indeed, at 9am our framework has the highest number of people arriving at work on time, followed by within-day re-planning Baseline, A2C, and CoEAs. Similarly, at 5pm our framework has the highest number of people arriving back at home, while under the other three policies most are either still at work or slowed by congestion on roads. Finally, analyzing the figure horizontally, the people in our framework spend the least time on roads (links 1, 6, 15, 20, 21, 22, and 23), and the most time doing useful activities at locations (home and work).

\begin{table}
\caption{Comparing our framework, CoEA, and A2C for \textit{average trip time} in minutes, \textit{on-time arrival ratio}, and \textit{expected reward} per vehicle per hour. Results are obtained using the SynthTown, Berlin, Dakar, and Santiago de Chile datasets.}
\label{table:comparison_results} %
\begin{tabular*}{1\columnwidth}{@{\extracolsep{\fill}}@{\extracolsep{\fill}}>{\raggedright}p{1.2cm}>{\raggedright}p{1.8cm}>{\raggedright}p{1.5cm}>{\raggedright}p{1.5cm}>{\raggedright}p{1.5cm}l}
\hline 
\textbf{Dataset}  & \textbf{Models}  & \textbf{Average trip time}  & \textbf{On-time arriving ratio}  & \textbf{Expected reward}  & \tabularnewline
\hline 
\multirow{4}{1.5cm}{\textbf{SynthTown}}  & Baseline  & 45.57  & 0.88  & 1.05  & \tabularnewline
& Our framework & \textbf{31.49}  & \textbf{0.89}  & \textbf{2.93}  & \tabularnewline
& CoEA  & 55.47  & 0.85  & -0.05  & \tabularnewline
& A2C  & 50.64  & 0.88  & 0.30  & \tabularnewline
\hline 
\multirow{4}{1.5cm}{\textbf{Berlin}}  & Baseline  & 39.65  & 0.72  & -20.65  & \tabularnewline
& Our framework & \textbf{38.38} & \textbf{0.86} & \textbf{-4.83}  & \tabularnewline
& CoEA  & 40.27  & 0.68  & -54.00  & \tabularnewline
\hline 
\multirow{4}{1.5cm}{\textbf{Dakar}}  & Baseline  & 29.13  & 0.86  & -5.27  & \tabularnewline
& Our framework & \textbf{28.12} & \textbf{0.90} & \textbf{0.78}  & \tabularnewline
& CoEA  & 30.30  & 0.85  & -10.06  & \tabularnewline
\hline 
\multirow{4}{1.5cm}{\textbf{Santiago}}  & Baseline  & 36.84  & 0.81  & -15.33  & \tabularnewline
& Our framework & \textbf{35.36} & \textbf{0.83} & \textbf{-7.58}  & \tabularnewline
& CoEA  & 38.30  & 0.75  & -20.79  & \tabularnewline
\hline 
\end{tabular*}
\end{table}

\textbf{Comparing summary performance metrics}: Table \ref{table:comparison_results}
compares average trip time, on-time arrival ratio, and average unit reward statistics of the four models using the SynthTown, Berlin, Dakar, and Santiago de Chile datasets. The Berlin, Dakar, and Santiago de Chile datasets are too large for the A2C model to run, demonstrating the superior scalability of our framework and CoEA. This comparison leads us to the same conclusion as the detailed comparison on the SynthTown data. Specifically, our framework has the lowest average trip time, the highest on-time arrival ratio, and highest expected reward among all datasets. The within-day re-planning algorithm works better than the co-evolutionary algorithm because the former uses instantaneous traffic information for trip planning in terms of average trip time, on-time arrival ratio, and the Charypar-Nagel score. Our algorithm based on solving a partially observable Markov decision process problem works better than the within-day re-planning algorithm, because we specifically optimize the Charypar-Nagel scoring function through traffic prediction and optimization within POMDP.

\section{Conclusions and Discussions}\label{conclusions}

The capability to sense the movement of millions of vehicles through mobile phones and to offer drivers trip plans has enabled a new way of controlling city traffic dynamics by turning transportation big data into insights and actions in a closed-loop. In this paper, we have presented a study showing that floating car data can indeed be utilized to offer optimal departure times and route choices associated with lower average trip time, higher on-time arrival ratio, and higher Charypar-Nagel score compared with how people normally travel, validated both in an agent-based transportation simulator and in the real world. The study is based on optimizing a partially observable discrete-time decision process and is evaluated in one synthesized scenario, one partly synthesized scenario, and three real-world scenarios. 

The results show that the framework of turning transportation big data into insights and actions in the real world is very much worth further research. For example, while the study in this paper is based on maximizing the Charypar-Nagel scoring function, further research is needed to identify the utilities of people in the real world in the framework of inverse reinforcement learning \cite{yang2020variational,yang2020bayesian}. While a particle filter seems to be sufficient for traffic-prediction and city traffic optimization in this study, further research is needed to lower the variance of Monte Carlo integration \cite{xu2016using,fang2017expectation}. In this paper we approximated the queuing-network traffic dynamics of MATSim with a discrete-event process, but further work to turn transportation simulators into a reinforcement learning environment \cite{khaidem2020optimizing} will be useful to facilitate this ``living lab'' style of research.

\bibliographystyle{IEEEtran}
\bibliography{reference}
\vspace{-.4in}
\begin{IEEEbiography}[{\includegraphics[width=1in,height=1.25in,clip,keepaspectratio]{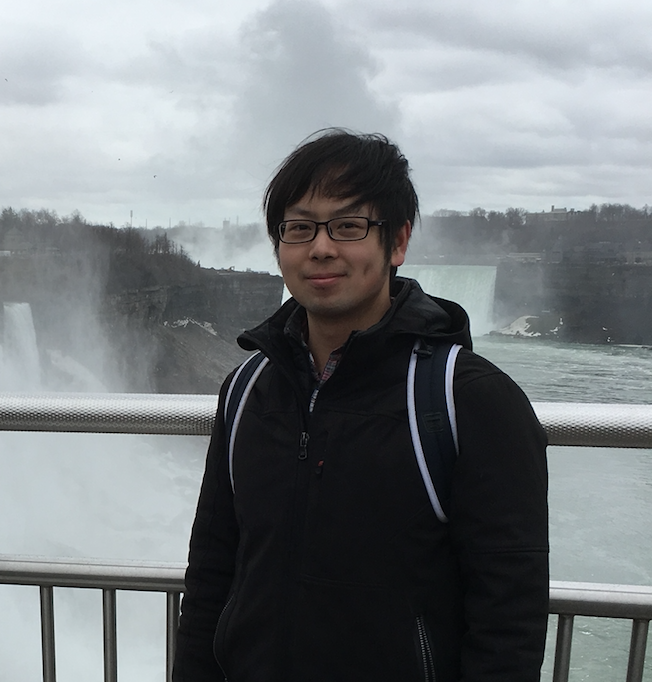}}]{Fan Yang}
is a PhD candidate in the Department of Computer Science and Engineering at the State University of New Your at Buffalo, USA. His research interests includes reinforcement learning, imitation learning, modeling, generative models, and probabilistic graphical models.
\end{IEEEbiography}\vspace{-.4in}
\begin{IEEEbiography}[{\includegraphics[width=1in,height=1.25in,clip,keepaspectratio]{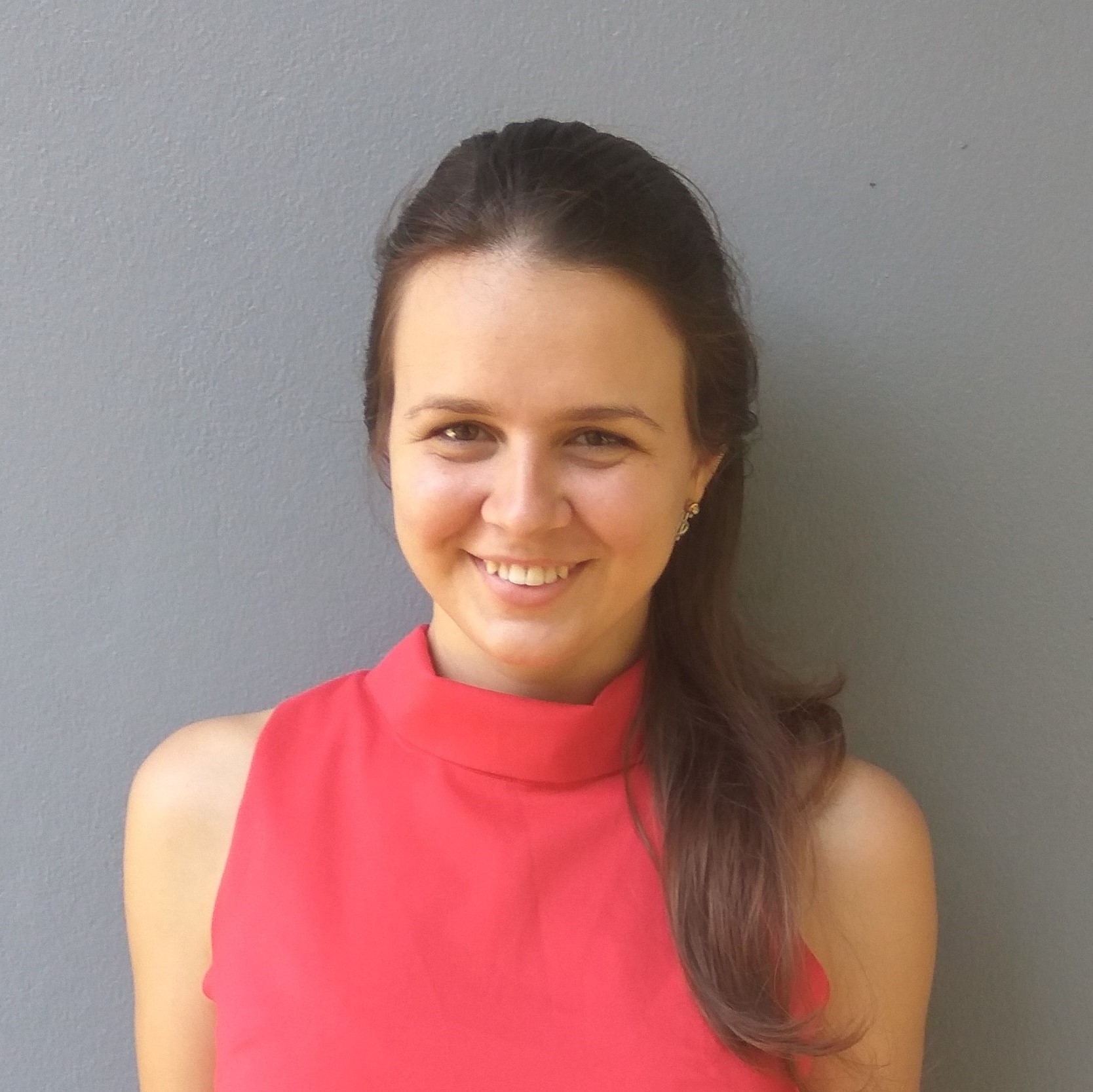}}]{Alina Vereshchaka}
is a PhD candidate in the Department of Computer Science and Engineering at the State University of New Your at Buffalo, USA. Her current research interests include deep reinforcement learning, optimization and multi-agent modeling in stochastic environments. She has conducted studies in the application areas of optimization, transportation and healthcare.
\end{IEEEbiography}\vspace{-.4in}
\begin{IEEEbiography}[{\includegraphics[width=1in,height=1.25in,clip,keepaspectratio]{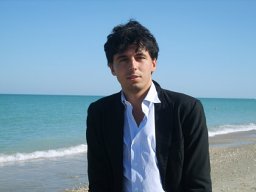}}]{Bruno Lepri}
received the Ph.D. degree in computer
science from the University of Trento, Italy, in 2009.
From 2010 to 2013, he was a joint Post-Doctoral
Fellow at MIT Media Lab, Boston, MA, USA, and
at Fondazione Bruno Kessler (FBK), Trento, Italy,
where he is currently leading the Mobile and Social
Computing Lab (MobS). He is also the Head of
the Research of Data-Pop Alliance, the first thinktank on Big Data and Development co-created by
Harvard Humanitarian Initiative, MIT Media Lab,
and Overseas Development Institute.
\end{IEEEbiography}\vspace{-.4in}
\begin{IEEEbiography}[{\includegraphics[width=1in,height=1.25in,clip,keepaspectratio]{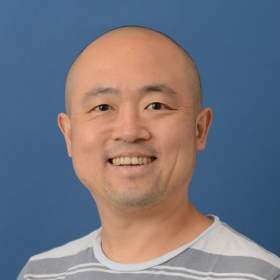}}]{Wen Dong}
is an Assistant Professor of Computer Science and Engineering with a joint appointment at the Institute of Sustainable Transportation and Logics at the State University of New York at Buffalo. His research focuses on developing machine learning and signal processing tools to study the dynamics of large social systems in vivo. He has a PhD degree from the M.I.T. Media Laboratory.
\end{IEEEbiography}
\vfill

\end{document}